\documentclass[aps, prb, amssymb, amsmath, 
twocolumn] {revtex4-2}

\usepackage{chngcntr}

\usepackage{graphicx}
\usepackage{color}
\usepackage[dvipsnames]{xcolor}
\usepackage{amsmath}
\usepackage{enumitem}
\usepackage{amssymb}
\usepackage[hidelinks]{hyperref}
\usepackage{cancel}
\usepackage{ulem}
\usepackage{multirow}

\newcommand{\be}{\begin{equation}}
	\newcommand{\ee}{\end{equation}}

\newcommand{\bea}{\begin{eqnarray}}
	\newcommand{\eea}{\end{eqnarray}}

\makeatother

\begin{document}

\preprint{APS/123-QED}

\setlength{\abovecaptionskip}{-60pt}

\title{Non-Analytic Magnetic Response and Intrinsic Ferromagnetic Clusters in a Dirac Spin Liquid Candidate}

\author{B.S. Shivaram$^{1,\dagger} $,  J. Prestigiacomo$^2$, Aini Xu$^4$, Zhenyuan Zeng$^{3,4}$, Trevor D. Ford$^1$, Itamar Kimchi$^5$, Shiliang Li$^{3,4,6}$, and Patrick A. Lee$^{7,\dagger} $}

\affiliation{} 
\affiliation{$^1$Department of Physics, University of Virginia, Charlottesville, VA. 22904, USA.}
\affiliation{$^2$Naval Research Laboratory, 4440 Overlook Drive, Washington D.C, USA.}
\affiliation{$^3$Beijing National Laboratory for Condensed Matter Physics,
Institute of Physics, Chinese Academy of Sciences, Beijing 100190, China.}
\affiliation{$^4$School of Physical Sciences, University of Chinese Academy of Sciences, Beijing 100190, China.}
\affiliation{$^5$School of Physics, Georgia Institute of Technology, Atlanta, Georgia 30332, USA.}
\affiliation{$^6$Songshan Lake Materials Laboratory, Dongguan, Guangdong 523808, China}
\affiliation{$^7$Department of Physics, Massachusetts Institute of Technology, Cambridge, MA.-02139, USA.}

\date{\today}

\begin{abstract}
Finding distinct signatures of a quantum spin liquid (QSL) is an ongoing quest in condensed matter physics, invariably complicated by the presence of disorder in real materials. In this regard the 2D Kagome system  YCu$_3$(OH)$_6$[(Cl$_x$Br$_{(1-x)}$)$_{3-y}$(OH)$_y$] (YCOB-Cl), where the vast mismatch in size of Y and Cu avoids subsitutional disorder, otherwise present in kagome materials, has emerged as a favorable candidate.  In crystals of this system, with $x<$ 0.4  and no long range order, we report an unusual field dependent magnetization $M(B)$, where $M/B$ changes linearly with $|B|$, the absolute value of the field, in contrast to the expected quadratic behavior.  Model calculations with a distribution of ferromagnetic (FM) clusters faithfully capture observed features suggesting such clusters to be intrinsic to real QSL materials. YCOB-Cl has a field enhanced $T^2$ heat capacity as expected for a Dirac QSL but lacks a linear $T$ behavior in the spin susceptibility. By demonstrating that FM clusters dominate the contribution to the susceptibility but not the  heat capacity, our work paves the way towards reconciling the apparent inconsistency with a Dirac QSL. \\

\end{abstract}

\pacs{Valid PACS appear here}

\maketitle

The search for quantum spin liquids (QSLs) in recent decades has encompassed a variety of material systems.  To ensure that no long range order develops down to the lowest temperatures, to stabilize a QSL, a general focus has been on those systems where frustration is predominant.  In this respect two dimensional or quasi-two dimensional kagome systems with magnetic atoms at the corners of triangles are ideal candidates to host novel quantum spin liquid ground states. Following this suggestion considerable effort has been devoted to studies on systems such as Herbertsmithite \cite{NormanRMP2016}. However, materials like Herbertsmithe suffer from the fact that Cu can replace some of the Zn ions which are outside the kagome planes. The high concentration of local moments thus created obscure the physics associated with the putative spin liquid \cite{MurayamaPRL2022}. Recently a number of groups have studied a new class of materials where Y takes the place of Zn. Due to the large difference in ionic radii, the substitution of Y by Cu does not take place and the local moments are largely suppressed. An example is the perfect quantum kagome antiferromagnet (AF) YCu$_3$(OH)$_6$Cl$_3$, which has a Heisenberg AF exchange of the order of 100 K. Unfortunately this compound  shows a rapid rise in the magnetic susceptibility at 12 K which has been interpreted as a possible magnetic transition \cite{ArhPRL2020}.  While this system is well described by the 2D nearest-neighbor kagome antiferromagnetic (KAFM) model where no magnetic transition is expected, the out-of-plane Dzyaloshinskii-Moriya (DM) interaction is significant in this material and could be the perturbation that causes  magnetic order, as predicted by theory\cite{ BernuPRB2020,Messio2010,Cepas2008,HeringPRB2017}. A partial replacement of Cl by O in the compound YCu$_3$(OH)$_6$O$_x$Cl$_{(3-x)}$) with x=1/3, does not exhibit any static magnetic order\cite{BarthelemyPRM2019}. This could be due to the role of the disorder introduced by the O substitution. Yet another related compound, Y-kapellasite - Y$_3$Cu$_9$(OH)$_{19}$Cl$_8$, has also been under intensive studies \cite{Puphal2017,BarthelemyPRM2019,ChatterjeePRB2023}. Here the unit cell has tripled and the kagome bonds distorted in an ordered way and the AF transition is suppressed to 2.1K. These results are in broad agreement with theoretical expectations \cite{Hering2022}.\par

In efforts to further suppress the magnetic order, recently a new class of compounds,  YCu$_3$(OH)$_6$Br$_2$[Br$_{1-y}$(OH)$_y$], which we will refer to as YCOB, where the  Cl in the "perfect" kagome compound has been replaced by a combination of Br and OH, have been discovered and shown not to order down to 50 mK \cite{ZengPRB2022}. The compound YCOB, can be made with varying $y$. Extensive studies of the heat capacity $C$ and linear magnetic susceptibility $\chi$ in high quality single crystals of this new system have been studied in detail for y=0.67 \cite{ZengPRB2022} and y=0.5 \cite{Liu2022}. An important finding is that $C/T$ shows a linear T dependence at low T. Furthermore $C/T$ increases with an applied magnetic field $H$, giving rise to a linear in $H$ term beyond a threshold field. These are behaviors expected for a Dirac spin liquid where fermionic spinons obey a Dirac spectrum. Neutron scattering also shows anomalous behavior which supports this picture\cite{Zengpreprintneutron}. On the other hand, the magnetic susceptibility, $\chi$, does not show the linear T behavior expected for the Dirac spin liquid. Instead it rises rapidly as T is decreased and may saturate at a finite value. This apparent inconsistency inspired us to examine the magnetization curve in this family of compounds in greater detail. This family is further expanded by partial substitution of Br by Cl, resulting in YCu$_3$(OH)$_6$[(Cl$_x$Br$_{(1-x)}$)$_{3-y}$(OH)$_y$], 
which we will refer to as YCOB-Cl.\\

We note that these compounds are not completely without disorder.  They are known to have disorder of a special kind\cite{Liu2022} where the OH replaces the Br sites randomly above and below the hexagons formed by the Cu, which causes the exchange interaction around the hexagon to alternate in strength. The same kind of bond alternation is found in the Y-kapellasite Y$_3$Cu$_9$(OH)$_{19}$Cl$_8$ mentioned earlier, except that in that case the alternation appears in a periodic manner, leading to a complicated AF ordering. It is not known how this bond disorder affects the properties of YCOB-Cl.  These questions further motivate the studies reported here.\\

\begin{figure}[h]
\includegraphics[width=0.45\textwidth]{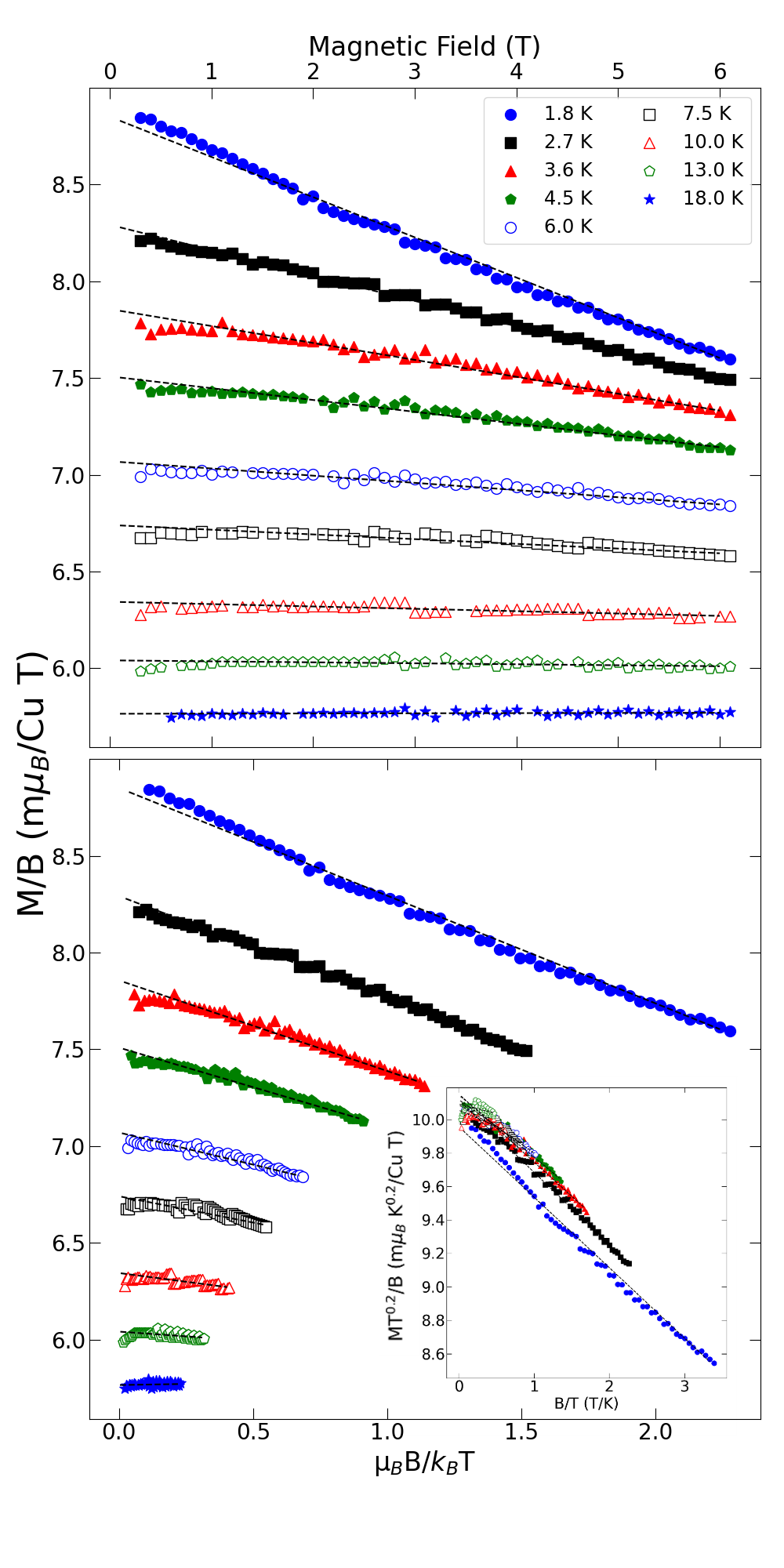}
\vspace{2.0cm}
\caption{\label{fig1} Magnetization $M/B$ isotherms obtained on the pure Bromine sample. Top: Shows $M/B$ vs $B$ at different temperatures as indicated. The slopes at high temperature are nearly zero but turn rapidly negative and diverge strongly as T$\rightarrow$0. Similar isotherms are obtained on the 21\% and 30\% chlorine doped crystals (see supplementary information).  We also verified that both field cooled (FC) and zero field cooled (ZFC) procedures gave identical isotherms. Bottom: $M/B$ plotted against a dimensionless ratio of magnetic field and temperature. Inset: Same data as in the bottom main panel to show that the curves collapse when scaled by $T^{0.2}$.}
\end{figure}

Generally speaking, disorder is expected to give rise to local moments with random coupling\cite{Bhatt1982}. This leads to the so called random singlet model (RSM), where the local moments are progressively paired up into singlets leaving behind fewer and fewer free spins as temperature is reduced. The main consequences of RSM is that $C/T$ and $\chi$ are predicted to go as $T^{-\gamma}$ where $\gamma$ is an exponent with a typical value of 0.2 to 0.5. Recently Kimchi, Sheckelton, McQueen and Lee \cite{KimchiNature2018} showed that the scaling exhibited in RSM (originally developed to explain behavior of isolated spins in Si:P) survives even with strong spin-orbit coupling and the presence of significant DM interactions.  In this model the magnetization is expected to scale as:

\begin{align}
M[B,T] \approx  B^{1-\gamma} F_q^M[T/B] 
\end{align}

For $B/T \ll 1$, the scaling function $F_q^M[T/B] \sim (T/B)^{-\gamma}$, and one expects the ratio $M/B$ to lowest order is field independent and to next order in $B/T$ is quadratic. This is strictly necessary for any finite system, such as a model of isolated singlet pairs, where there can be no singularity. Thus, there is a $B^2$ dependent perturbation to the ratio, $M/B$.  In the following we present our surprising experimental finding of a leading linear $|B|$ field dependence that extends to $|B|/T\ll 1$. This non-analytic behavior, of course, cannot continue to zero field, but our finding is that it extends to a surprisingly small $|B|/T$ ratio. This signals the failure of the random singlet model. The failure can be seen in $M/B$ (plotted against both the experimental field,  and the reduced magnetic field) being predominantly linear as presented in Fig.~\ref{fig1}.  The inset of Fig.~\ref{fig1}, bottom panel, shows the negative linear scaling produced with a small $\gamma=0.2$.  The linear rather than the expected quadratic behavior is predominant. A second case for the failure of RSM comes from the observation that $C/T$ goes to zero as $T$, instead of diverging as $T^{-\gamma}$ prescribed by the model. Thus we conclude that the randomness in YCOB-Cl is rather different from that assumed in RSM. It is quite possible that the bond randomness in YCOB-Cl is very different from the irregular arrangement of local moments due to the substitution of Zn with Cu one finds in Herbertsmithite where RSM seems to work. This failure of the random singlet model requires an alternate approach. Addressing a comprehensive set of magnetization measurements on a range of samples with different concentrations of Cl in the present work enables us to  propose such an approach, phenomenologically, involving clusters of ferromagnetically aligned spins.  \\

In our experiments we use single crystals of YCOB-Cl, 
YCu$_3$(OH)$_6$[(Cl$_x$Br$_{(1-x)}$)$_{3-y}$(OH)$_y$,
synthesized by a hydrothermal method similar to those reported previously \cite{ZengPRB2022}.  In this method the ratio between Cl and Br can be continuously tuned and we obtain hexagonal plates with the in-plane diameter up to 1.3 mm and the Cl doped samples being up to 1 mm thick. All the crystals were further ultrasonically cleaned in water to remove possible impurities attached to the surfaces of the crystals.  Magnetization measurements were carried out on a Quantum Design magnetic property measurement system employing a sample holder traversing both halves of the static SQUID detection coils to eliminate the background contribution to the measured moment.  No hysteresis was observed within the field range of the data reported in this paper (B $>$0.1 T) on all samples (pure Br, 21\% Cl, and 30\% Cl). \\

\begin{figure}
\includegraphics[width=0.5\textwidth]{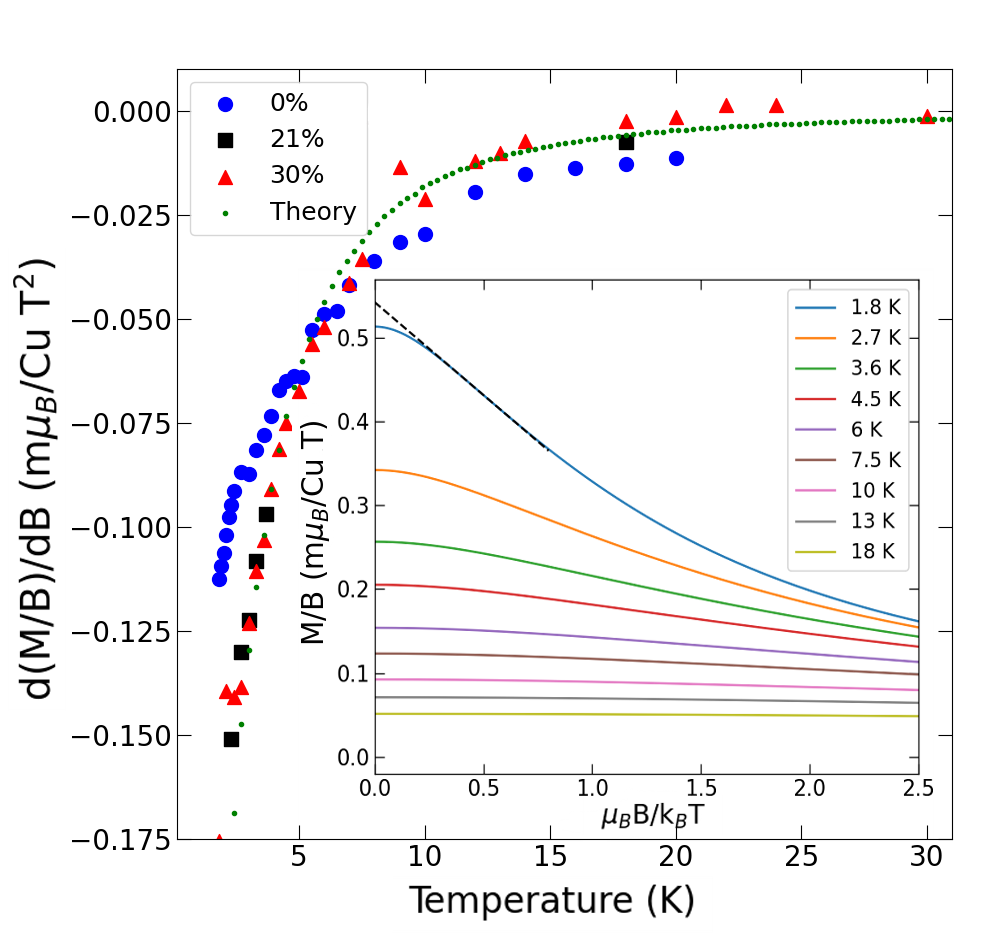}
\vspace{1.7cm}
\caption{\label{fig2}Shows the slopes of the linear region in plots as in fig.1 as a function of temperature for all three compositions of chlorine 0 \% (green circles), 21\% (black circles) and 30\% (red circles).  The green circles are from a theoretical model of adding the magnetization of free spins with moments up to $S=4$ as explained in detail in the text. The inset shows the extended linear region of the isotherms used to extract the slopes (green circles main panel) in the model. }
\end{figure}

In fig.1 we show the magnetization isotherms measured on the pure Br sample plotted as $M/B$ against the field $B$, oriented parallel to the hexagonal plane, so that we can remove the linear term and highlight the nonlinear response. Such plots enabling separation of the nonlinear part from the initial linear magnetization in paramagnetic systems or in the non-ordered state are normally displayed as $M/B$ vs. $B^2$. The slope of the resulting plots provide the third order susceptibility \cite{HolleisNPJ2021, ShivaramPRB2018, ShivaramPRB}.  The occurance of a negative linear term in plots of M/B vs. B down to the lowest fields as seen here is very unusual and unique and suggests a non-analytic behavior of the magnetization.  It is also clear that the negative slope in fig.1 increases rapidly towards low temperature. The slopes obtained from linear fits to these isotherms are shown as a function of temperature in fig.2.  In addition to the pure Br sample measurements of isotherms were also carried out on samples doped with 21\% and 30\% Cl replacing the Br.  These samples also exhibited the same behavior and fig.2 displays the results obtained on the slopes of the $M/B$ vs $B$ isotherms on these samples as well. The remarkable similarity of the nonlinear contribution irrespective of the Cl doping attests to the robust nature of this unusual quadratic response and also implies that the same type of disorder is at play irrespective of Cl doping.\\

\begin{figure}
\includegraphics[width=0.5\textwidth]{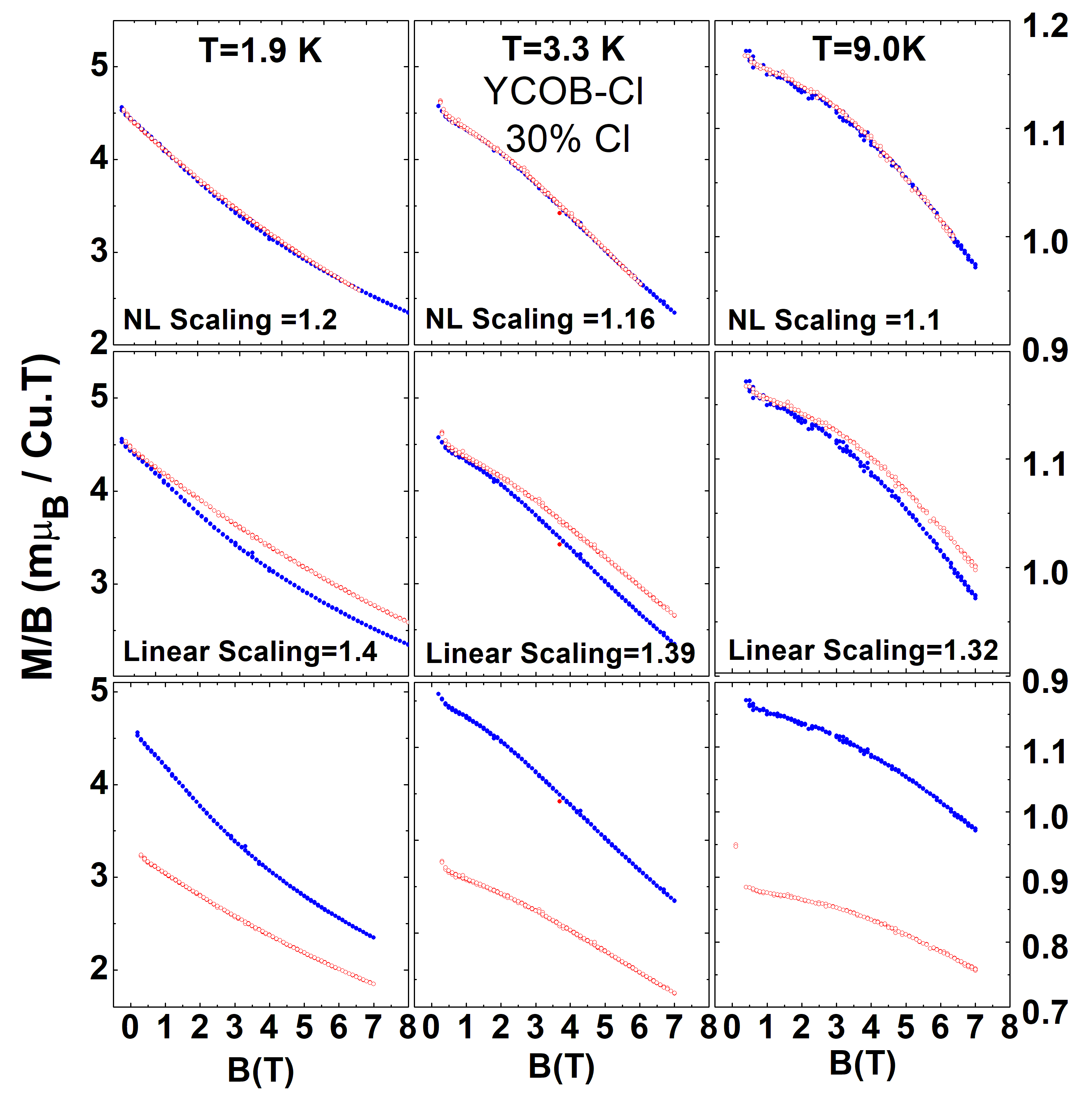}
\vspace{1.8cm}
\caption{\label{fig3}Shows the anisotropy of the linear and nonlinear magnetization responses of the 30\% Cl sample. Results for three separate temperatures are shown each arranged in vertical panels. The bottom panel shows the unscaled raw data for $B\parallel$c-axis (blue) and $B\perp$c-axis (red).  In the middle panel the $B\perp$c-axis data is scaled up to match the B=0 values (linear susceptibility scaling).  In the top panel the nonlinear susceptibilities merge when the B axis for $B\perp$c-axis is scaled by the NL Scaling factor shown. For T=3.3K the vertical scale begins as shown on left but is expanded by a factor of two. The scale on the right applies only to the 9.0K data.
}
\end{figure}

In fig. 2 we also present the slope of the linear region of plots of $M/B$ vs. $B$ obtained through model calculations  with ferromagnetic clusters. In these calculations the magnetization is given by a sum over small cluster contributions:

\begin{align}M=\sum_{2S=1}^{8} n\left(S\right)g\mu_BSB_S(x)\end{align}
where $S$ is the spin of the clusters.  $B_S(x)$ is the Brillouin function, $n(S)$ is the weighting factor for a cluster with spin $S$, and the parameter $x$ is
\begin{align}x=\frac{Sg\mu_BB}{k_BT}\end{align}

The density $n(S)$ of each spin cluster was adjusted to produce the largest linear region in $M/B$ vs, $B$.  This maximization was attempted empirically using simple monotonically decreasing  functions $n$ of the form $n(S) = S^{-a}$ ( with $a$ a natural number) to represent the weighting of each possible spin cluster up to $S=4$. A value of $a=3$ provided reasonable model results that closely followed the data as shown by the green dots in fig.2. In the supplementary information document we provide an alternate figure with $a=2$ for comparison.\\

    While Fig.2 shows that the model successfully captures the size as well as the temperature dependence of the B linear slope in $M/B$, a closer examination of the results shown in the inset and its comparison with the data shown in the lower section of Fig 1 reveals two inadequacies. First the linear behavior of the model calculation extends only to $\mu_B /kT$ of about 0.25 whereas the data extend down to about 0.1. Second, while the theory does not include the background contribution from the spins outside the cluster and the absolute value of $M/B$ cannot be directly compared, we should be able to compare the diference in $M/B$ between the lowest and a relatively high temperature, such as 10 K. For this difference the theory gives about 0.5 m$\mu_B$/Cu-T while the data is about 2.5 m$\mu_B$/Cu-T, about a factor of 5 larger.  For the first point we note that in our model the usual $B^2$ correction to $M/B$ changes over to linear $B$ at a scale given by $\mu_B/kT \approx 1/S$. Thus the linear region can be extended further by assuming clusters of $S > 4$. These large spin clusters will also increase the size of the temperature dependence of $M/B$ and can help improve the disagreement in the amplitude noted for the second point. However, in order to include larger spins while maintaining the good agreement of the linear slope shown in Fig 2, it will be necessary to introduce more parameters to characterize the spin distribution. We did not pursue such elaborate modelling and prefer to stay with a simple power law distribution to make the point that FM clusters are the primary  cause to account for the data.\\

As a further test of the above model we explore the behavior of the nonlinear term with the orientation of the magnetic field with respect to the crystal axes. In fig.3 we show the isotherms M/B vs. B for both orientations of the magnetic field at three different temperatures.  A pair of isotherms (for the two orthogonal orientations, B$\parallel$c-axis and B$\perp$c-axis) at a common temperature are displayed stacked in a set of three vertical panels.  The bottom-most panel displays the raw data while in the middle panel the isotherm for B$\perp$c-axis has been scaled up by the numerical factor 'linear scaling' as given in the figure.  This scaling of the linear susceptibility ensures that the M/B values at B=0 match.  Next, in the top most panel we display the same isotherms but this time we scale the magnetic field of the B$\perp$c-axis data so as to collapse the two curves.  This gives the nonlinear susceptibility scaling factor, as dislayed in the figure. The anisotropy of the linear susceptibility is proportional to the square of the g-factor anisotropy as per eqn (2).  Further, since the Brillouin function involves only the product g.B, in order to preserve the scaling of the nonlinear susceptibility the magnetic field has to be changed by the square root of the linear susceptibility scaling factor.  We tested for this in the experimental data and the results for the 21\% and 30\% chlorine samples are shown in fig.4.  The square root scaling is perfectly obeyed.  In performing the above anisotropy scaling analysis we subtracted the field independent M/B values at 25 K as a high temperature background not arising from the FM clusters.\\

\begin{figure}
\includegraphics[width=0.5\textwidth]{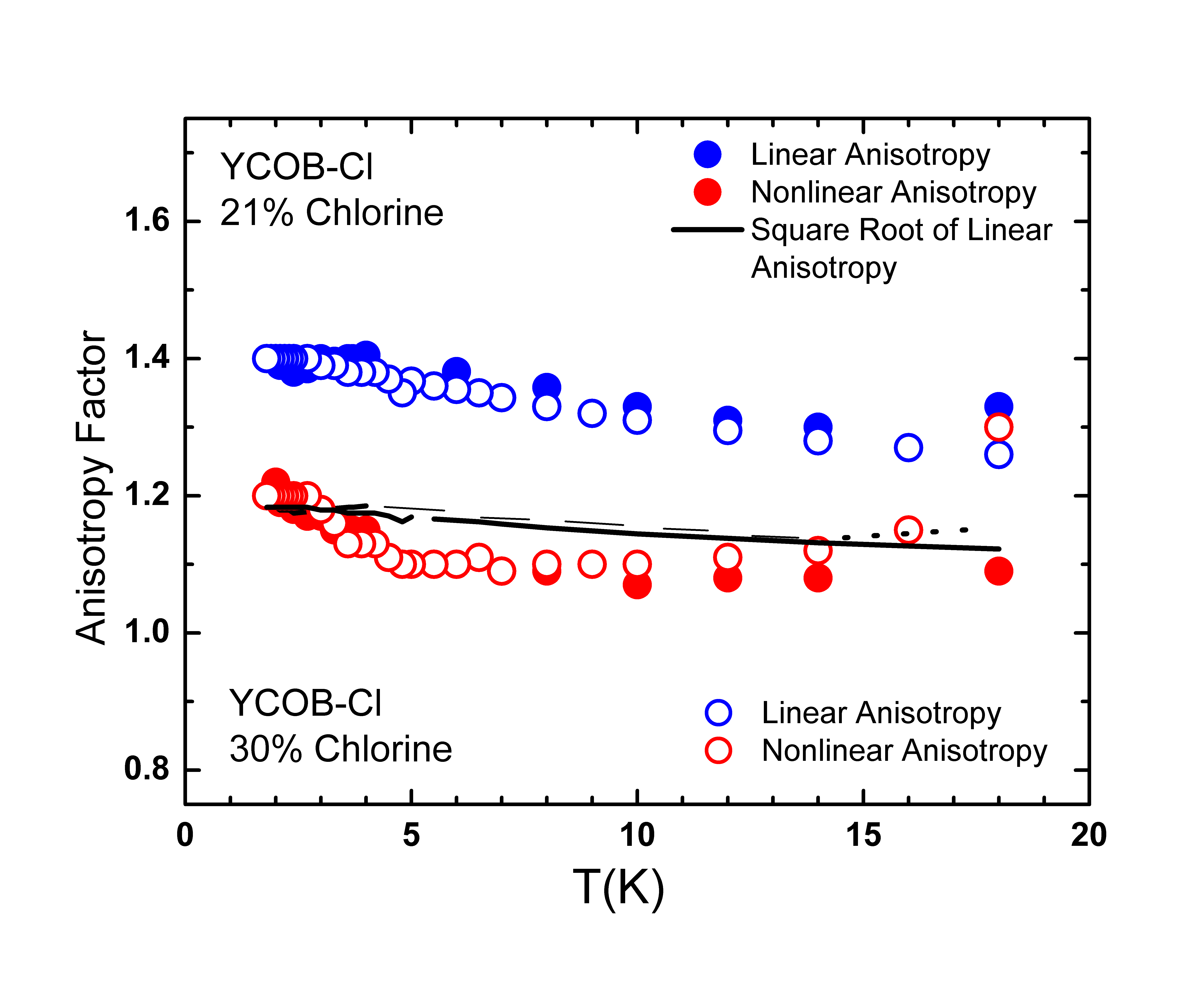}
\vspace{1.5 cm}
\caption{\label{fig4}Shows the linear scaling and nonlinear scaling factors as obtained from the procedure shown in fig.3. Results for samples with two different Cl concentrations, 21\% and 30\%, are shown.  The nonlinear scaling factor is reduced by a square root factor from the linear one as expected from the FM cluster model.
}
\end{figure}

\textbf{Discussion}: The identification of the salient experimental features of a quantum spin liquid state is an ongoing effort.  In the above we have presented comprehensive magnetization results on a new system, YCu$_3$(OH)$_6$[(Cl$_x$Br$_{(1-x)}$)$_{3-y}$(OH)$_y$, in which Cl replaces Br. This substitution as well as the presence of the (OH)$^-$ ions necessarily introduces disordered spins which could cluster. These clusters make their own imprints on the measured magnetic response. Indeed in our experiments on samples with $x < 0.4$ we find that the leading order correction  to the magnetization divided by $B$ is linear in $|B|$. This unusual non-analytic behavior of the nonlinear susceptibility, which extends to low magnetic fields $\approx 0.1$ T,  appears to be concomitant with an ubrupt or sharp departure of the linear $\chi$ from a Curie-Weiss behavior seen in many candidate QSL systems.  In these systems typically below the temperature range 20-25 K $\chi$ starts to rise rapidly signalling an enhanced FM behavior.   \\

Indeed, in analysis of our data we develop a model based on a distribution of FM clusters. These clusters dominate the contribution to spin susceptibility but contribute only a small amount to the specific heat. This observation paves the way towards resolving the puzzle of why the Dirac like response in the specific heat and its magnetic field dependence in YCOB appear to be inconsistent with the spin susceptibility data and potentially removes a discrepancy in the interpretation of this system as Dirac SL.  Since disorder driven formation of FM clusters in strongly frustrated magnets can be widespread, the insights gained from this work should asisst in a better understanding of other QSL candidate materials as well \cite{LiSR2015, KhatuaNatureComm2022}. Our work may also be relevant to cold-atom lattice experiments simulating frustrated many body Hamiltonians \cite{ChenPRA2010, XuNature2023}. \\

\textbf{Acknowledgements:} The work at the University of Virginia (BSS and TDF) was supported by NSF Award DMR-2016909.  PL acknowledges support by DOE (USA) office of Basic Sciences Grant No. DE-FG02-03ER46076. IK acknowledges support by Grants NSF PHY-1748958 and PHY-2309135 to the Kavli Institute for Theoretical Physics and PHY-1607611 to the Aspen Center for Physics. The work by AX, ZZ and SL is supported by the National Key Research and Development Program of China (Grants No. 2022YFA1403400, No. 2021YFA1400401), the Strategic Priority Research Program(B) of the Chinese Academy of Sciences (Grants No. XDB33000000, No. GJTD-2020-01). The work at the Naval Research Laboratory (JP) has been funded by the Office of Naval Research, under the NRL 6.1 Base Program. \\

$^\dagger$Corresponding authors:\par
\hspace{\parindent}    bss2d@virginia.edu\par
\hspace{\parindent}   palee@mit.edu\\

\bibliography{ref}
\end{document}